\documentclass[nofootinbib, floatfix,longbibliography,twocolumn]{revtex4-1}
\usepackage{amsmath}
\usepackage{amssymb}
\usepackage{graphicx}
\usepackage{verbatim}
\usepackage{color}

\AtBeginDocument{}
\usepackage[pdftex, breaklinks, colorlinks, citecolor=blue, urlcolor=blue]{hyperref}

\begin{document}

\author{A. M. Jones, E. J. Pritchett, E. H. Chen, T. E. Keating, R. W. Andrews, J. Z. Blumoff, L. A. De Lorenzo, K. Eng, S. D. Ha,  A. A. Kiselev, S. M. Meenehan, S. T. Merkel, J. A. Wright,  L. F. Edge, R. S. Ross, M. T. Rakher, M. G. Borselli, A. Hunter} 
\affiliation{HRL Laboratories, LLC, 3011 Malibu Canyon Road, Malibu, California 90265, USA}

\begin{abstract}

We implement a technique for measuring the singlet-triplet energy splitting  responsible for spin-to-charge conversion in semiconductor quantum dots.   This method, which requires fast, single-shot charge measurement, reliably extracts an energy in the limits of both large and small splittings.
We perform this technique on an undoped, accumulation-mode Si/SiGe triple-quantum dot and find that the measured splitting varies smoothly as a function of confinement gate biases.  Not only does this demonstration prove the value of having an {\it in situ} excited-state measurement technique as part of a standard tune-up procedure, it also suggests that in typical Si/SiGe quantum dot devices, spin-blockade can be limited by lateral orbital excitation energy rather than valley splitting.

\end{abstract}

\title{
Spin-Blockade Spectroscopy of Si/SiGe Quantum Dots
}
\maketitle

Qubits based on electrons confined in silicon quantum dots (QDs) benefit from the advantages inherent to many semiconductor platforms: fast control, small form factors, and established fabrication methods. Moreover, mitigation of a dominant decoherence pathway---interactions with non spin-zero nuclei---is possible with isotopic enhancement \cite{LADD2018}. These advantages have led to key demonstrations, including extended coherence \cite{Muhonen2014,eng2015,Yoneda2018} as well as fast, accurate control via a number of techniques including tunnel barrier modulation \cite{reed2016,martins2016}, electron spin resonance \cite{Veldhorst2014,Muhonen2015,Fogarty2017}, and induced spin-orbit coupling \cite{kawakami2016}.  Recently, one-qubit  randomized benchmarking \cite{Veldhorst2014,Zajac2017,Yang2018}, two-qubit randomized benchmarking \cite{Huang2018}, and two-qubit entangling sequences \cite{Veldhorst2015,Zajac2017,Watson2018} have been demonstrated with Si-based qubits.

Many quantum dot based qubits rely on the conversion of electron spin states, which preserve quantum information by interacting weakly with their electrical environment, to charge states for control and measurement.  In exchange-based schemes, distinct spin states of two interacting electrons (spin-singlets and spin-triplets) are converted into distinguishable charge states utilizing the well-established principles of Pauli spin blockade (SB)\cite{Hanson2007}. 
The robustness of SB relies on the energy separation between the ground-singlet and ground-triplet levels of two electrons occupying the same dot, denoted here as $\Delta_{\rm SB}$. This energy  limits state measurement and preparation fidelities by setting the range over which one can achieve spin-to-charge conversion and the accuracy with which one can initialize a ground state singlet in a system at finite temperature  \cite{Gamble2016}.

\begin{figure}[h]
\includegraphics[width=1\linewidth]{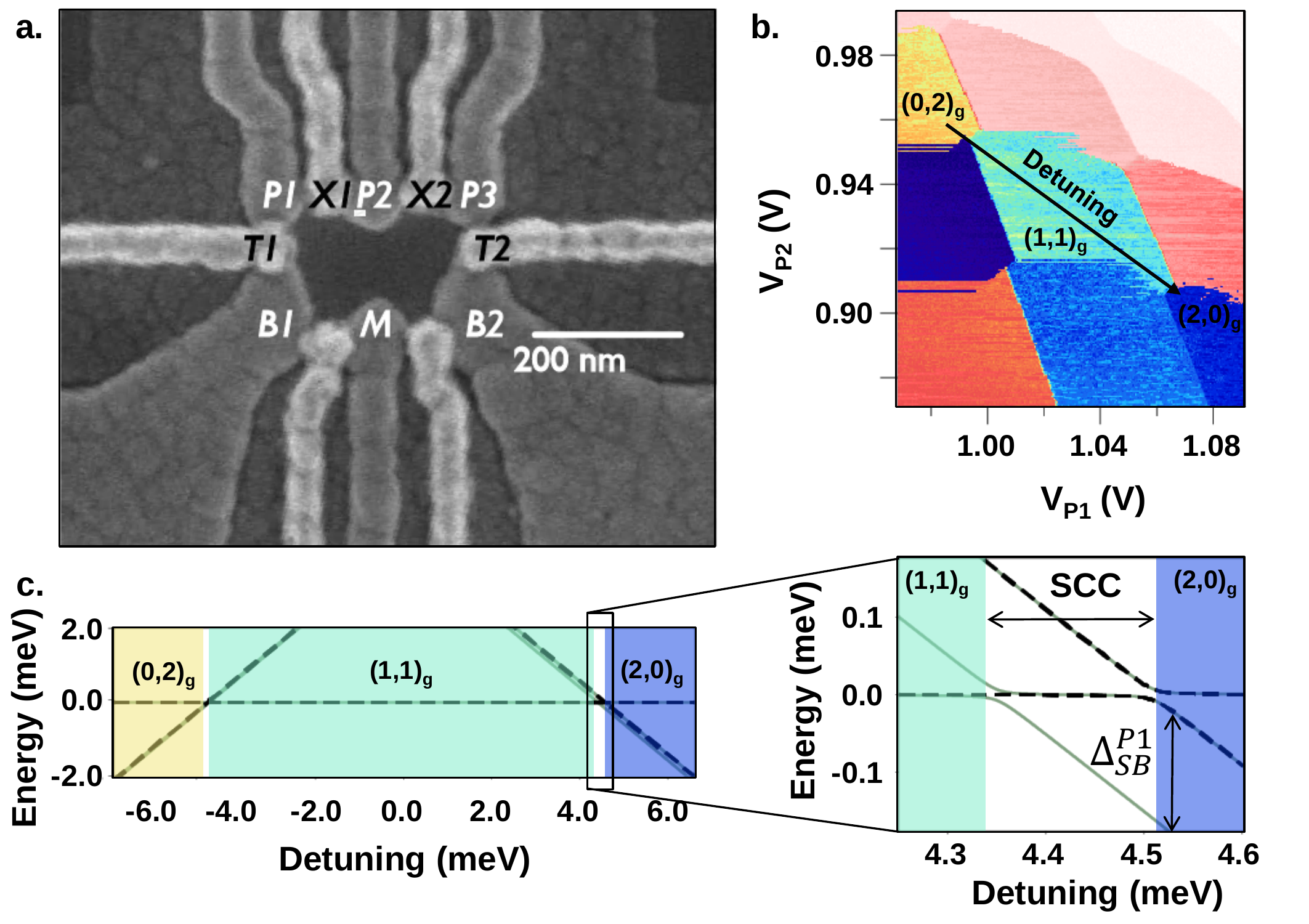}
\caption{Device scanning electron micrograph, charge stability, and energy level schematic.  {\bf{a}}, Scanning electron micrograph of a triple-dot device defined by gates P1, P2, P3, with nearby quantum dot electrometer, M.  {\bf{b}}, Charge stability map vs. applied voltage on gates P1 and P2, with color indicating measured current through charge sensor M. {\bf{c}}, Energy levels vs. detuning of a two-electron Fermi-Hubbard model for spin singlet (solid) and triplet (dashed) states.}
\label{fig1}
\end{figure}

Establishing what sets $\Delta_{\rm SB}$ in Si QDs is crucial for qubit design.
 While orbital excitation energies are, in principle, easily manipulated by changing confining potentials with gate biases, the maximum attainable $\Delta_{\rm SB}$ is conventionally thought to be limited by valley splitting,  the energy splitting between the two lowest-lying quantized states originating from conduction band valleys  of bulk, tensily strained silicon \cite{Boykin2004}.  Two-electron orbital energies are determined by the strength and symmetry of in-plane confinement  \cite{Melnikov2006,Hanson2007}, whereas valley splittings are determined primarily by the details of out-of-plane electron confinement, a challenge that has inspired many theoretical and engineering efforts \cite{Boykin2004,Sasaki2009,Zhang2013,Neyens2018}.  Numerous techniques have been used to estimate the magnitude of valley splitting in Si/SiGe quantum dots,  including photon-assisted tunneling \cite{VanDerWiel2002}, cavity coupling \cite{Mi2017}, Landau-Zener-St\"{u}ckelberg interferometry \cite{Schoenfield2017}, magnetotransport \cite{Neyens2018}, and  magnetospectroscopy \cite{Lim2011,Borselli2011}.  
However, the results of these techniques may not accurately measure $\Delta_{\rm SB}$ as they probe length scales not generally relevant to QD electrons, require the application of additional electromagnetic fields, or use biasing configurations different from nominal operation.
These conditions impede a direct translation of the measured energies to qubit performance.

In this Letter, we describe an excited-state spectroscopy technique which directly measures $\Delta_{\rm SB}$ in QDs. 
The utility of this method lies in its ability to extract both small and large $\Delta_{\rm SB}$ in exactly the bias configuration required to achieve spin-to-charge conversion and  without relying on a detailed physical model or the addition of  control fields beyond those normally required to operate exchange-only quantum dot qubits.   
We discuss the requirements placed on our measurement system to perform this technique, which is fundamentally limited by the fidelity of single-shot state discrimination.   
We then demonstrate this technique on a Si/SiGe double-dot and find $\Delta_{\rm SB}$ to be highly tunable and smoothly varying with bias voltages, suggesting that $\Delta_{\rm SB}$ in this device is limited by orbital confinement energy rather than valley splitting \cite{Friesen2010,Shi2011,Yang2013}.
Both the establishment of a reliable method for {\it in situ} excited-state spectroscopy and the identification of limitations to device performance should be relevant  to many QD architectures.

We measure an accumulation-mode, Si/SiGe triple dot (Fig.~\ref{fig1}a) which is similar to other devices \cite{Lai2011,Borselli2015,Zajac2015,Ward2016,Studenikin2018} that have demonstrated both a scalable platform for trapping many electrons   \cite{Zajac2016} and universal qubit control \cite{eng2015,reed2016}.   
Building on our previous experience \cite{Borselli2015,deelman2016}, electrons are  confined in a tensile-strained Si quantum well, embedded in a strain-relaxed, undoped SiGe alloy. Lateral confinement is provided by two types of overlapping metal gates: large field gates that prevent the accumulation of electrons in undesirable locations and smaller control gates.  
Quantum dots are defined under the ``plunger" gates  labeled $P1$, $P2$, and $P3$ and are loaded from electron reservoirs under the ``bath" gates $B1$ and $B2$.  
Tunneling from the baths to the dots is mediated by the ``tunnel" gates $T1$ and $T2$, and interdot tunnel coupling is controlled by the ``exchange" gates $X1$ and $X2$. 
 Bound electrons capacitively couple to the ``measure" dot $M$, 
 with each charge configuration affecting the measured conductance as seen in Fig.~\ref{fig1}b.
 Gate biases provide control of the triple dot through two mechanisms: a linear response in the chemical potentials, which can sensitively calibrate away any residual charge disorder, and an exponential response in tunnel coupling, which allows fast control with large on/off ratios and ultimately facilitates high performance quantum gates.

\begin{figure}[h]
\includegraphics[width=1\linewidth]{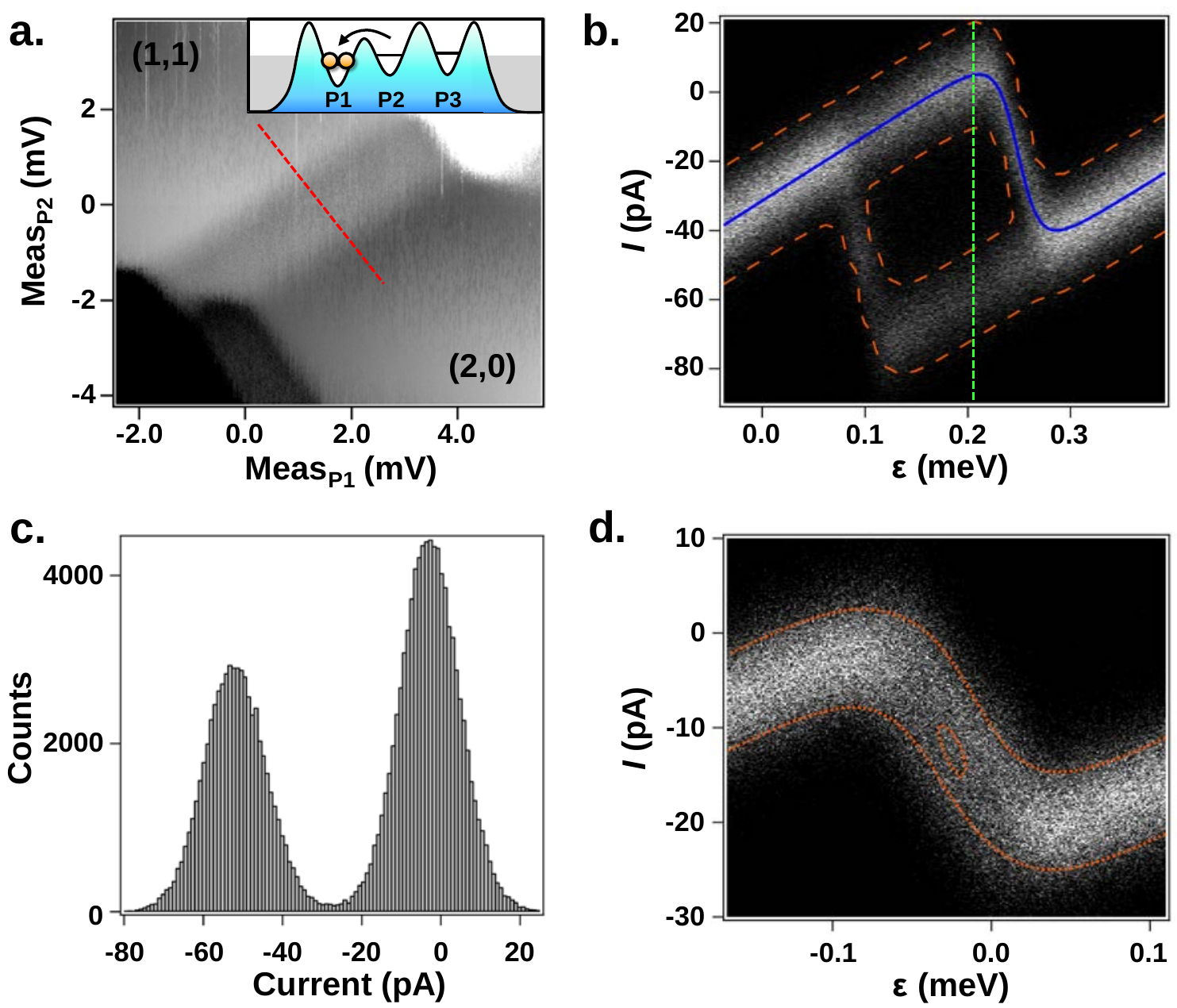}
\caption{Single-shot measurements of two-electron ground- and excited-state avoided crossings.  {\bf{a}}, Charge stability scan around the (2,0)-(1,1) avoided crossing, employing a prepare-dephase-measure pulse sequence. Grayscale contrast derives from differential measurement of current through charge sensor M. Inset: Cartoon depicting 1D potential profile of dots $P_1$, $P_2$, and $P_3$ for biasing conditions corresponding to the center of the main figure. Moving away from (1,1) towards (2,0) dumps both electrons (yellow circles) into the dot $P_1$.  {\bf{b}}, 5000 single-shot measurements taken at each point along the red detuning line in {\bf{a}}. The vertical axis shows the measured current value, with the horizontal axis giving detuning, in meV, from a point near the (2,0)-(1,1) anticrossing. Grayscale contrast corresponds to the number of counts at each detuning/binned current value. Orange dashed lines show contour line from fit to  Eq.~(\ref{eq:FermiHubbard3D}). Blue line shows the triplet-branch profile from Eq.~(\ref{eq:FermiHubbarddiCarlo}). {\bf{c}}, 150,000 shot histogram taken at the detuning indicated by the green line in {\bf{b}}.  {\bf{d}}, data from the (2,0)-(1,1) anticrossing of a second similar device. 500 shots were taken at each detuning. Orange dashed lines show contour line from fit to  Eq.~(\ref{eq:FermiHubbard3D}).}
\label{fig2}
\end{figure}

Spin-to-charge conversion (SCC), by which two-electron spin states are  converted into distinct charge states, occurs at biases where two electron number-conserving charge states are nearly degenerate, e.g., bottom-right: (2,0)-(1,1) and top-left: (0,2)-(1,1) of Fig.~\ref{fig1}b \cite{Kane1998,Hanson2007}.  As dictated by the Pauli exclusion principle, the two-electron spin-symmetric triplet must be antisymmetric in some other degree of freedom.   In silicon, this can be provided by valley, orbital excitation, or some hybridization of the two \cite{Friesen2010}.
The ground-singlet to ground-triplet energy splitting of two electrons occupying dot ${P}i$, denoted $\Delta_{\rm SB}^{Pi}$, determines the range of detunings that support SCC, i.e., biases at which the ground-triplet remains in the (1,1) charge state while the ground-singlet occupies a single QD  (see Fig.\ref{fig1}c).
These distinct charge states result in different electrostatic potentials at the measure dot, leading to measurably different conductances at this charge sensor \cite{Borselli2015}.

Our technique measures $\Delta_{\rm SB}^{Pi}$ by 
analyzing single-shot current values as singlet/triplet mixed states are swept adiabatically
through detunings that support spin to charge conversion. This involves a four-step pulse sequence: (1) state preparation, (2) spin dephasing, (3) measurement near SCC, and (4) a final charge reference measurement. In step (1) we prepare a spin-singlet by biasing near the (1,0)-(2,0) charge boundary where fast cotunneling processes quickly relax (2,0) triplets to (2,0) singlets.   
For step (2), to prepare a mixed state, we adiabatically traverse the (2,0)-(1,1) anticrossing and rapidly pulse to the center of the (1,1) cell.  
Then the spin-singlet is allowed to dephase in a bath of fluctuating non spin-zero nuclei for 10 $\mu$s, which is long compared to the measured $T_2^*\simeq2~\mu$s in this 800 ppm $^{29}$Si device. In step (3) the resulting mixed state is rapidly biased back near the (1,1)-(2,0) anticrossing, then adiabatically ramped to a measurement point, which is swept through detunings near and including SCC.  A second measurement (4) is performed deep within the (2,0) cell  and subtracted from the first to remove the effects of low-frequency charge noise.   This sequence is very similar to those used for standard quantum measurements when operating the device as a singlet-triplet qubit. However, high-fidelity state preparation in step (1) is unnecessary as the remaining steps work identically for a mixture of ground singlets and triplets; therefore this step could be skipped altogether if care is taken that the electrons are not excited out of the singlet/triplet ground state manifold.

The region of SCC is revealed as a step in the contrast of conductance between measurements at steps (2) and (4), apparent in a two-dimensional  sweep of the biases on P1 and P2 during the first measurement  (see Fig.~\ref{fig2}a).  We extract the energy $\Delta_{\rm SB}$ by taking a line cut (marked in red in Fig.~\ref{fig2}a) perpendicular to the region of SCC and repeating the pulse sequence many times at each point along this detuning axis, recording the result of each repetition.  This results in a distribution of independently measured currents at each value of detuning.   We bin the number of counts at a given detuning within a range of measured current values to obtain the grayscale intensity plot of Fig.~\ref{fig2}b, revealing two distinct branches of the transition from the higher-current, (1,1) charge state to the lower-current, (2,0) charge state. With positive detuning defined to be in the (1,1) to (2,0) direction, we ascribe the charge transition that occurs at smaller detuning to a spin singlet state. The transition occurring at larger detuning then corresponds to a spin triplet, separated from the singlet branch by the lowest excited-state energy of both electrons occupying dot P1, $\Delta_{\rm SB}^{\rm P1}$.  The pulse sequence can easily be modified to sample detunings near the (0,2)-(1,1) anticrossing, with the second charge reference measurement performed deep in the (0,2) charge cell; the resulting data are qualitatively similar to that shown in Fig. \ref{fig2}a, but the extracted energy is $\Delta_{\rm SB}^{\rm P2}$.
Therefore, with only a slight addition to the pulse sequence, we are able to efficiently measure the excited state energies of two neighboring dots at a given bias configuration.
In order to accurately convert the applied detuning bias voltages to an energy scale, we incorporate both the gate lever arm strength and cross-capacitance between dots $P_1$ and $P_2$ in calculating a scaling factor. 
The uncertainty in our measurement of the lever arm constitutes a significant source of uncertainty in our measurement of $\Delta_{\rm SB}$ (see Supplemental Section I).

In order to standardize the extraction of $\Delta_{\rm SB}$, we fit
the intensity $z$ measured as a function of  binned current and detuning (Figure~\ref{fig2}b) to the following functional form:
\begin{eqnarray} \label{eq:FermiHubbard3D}
z(\Delta_{\rm SB}) &=& P_{\rm S}\cdot f(\epsilon-\epsilon^0,I-(I^0+m\epsilon),t_{\rm c}^{\rm S})\\
 &+& P_{\rm T}\cdot f(\epsilon-\epsilon^0-\Delta_{\rm SB},I-(I^0+m\epsilon),t_{\rm c}^{\rm T})\nonumber
\end{eqnarray}
where $P_{\rm S}$ and $P_{\rm T}$ are the relative populations of ground-singlet and ground-triplet with tunnel couplings $t_{\rm c}^{\rm S}$ and $t_{\rm c}^{\rm T}$, respectively. $\epsilon$ is the dot-to-dot detuning in units of energy, and $\epsilon^0$ is the detuning energy of the spin-singlet charge transition. $I$ is the differential current through dot M, with constant offset $I^0$ and linear offset $m\epsilon$; the slope $m$ stems from a linear cross-capacitance between dots $P_1/P_2$ and dot M.  The function
\begin{equation}
f(\epsilon,I,t_{\rm c})={\rm exp}\left[-\frac{1}{2 \sigma_I^2}\left({I-g(\epsilon,t_{\rm c})}\right)^2\right]
 \label{eq:FermiHubbardEachBranch}
\end{equation}
describes the current broadening by an amount $ \sigma_I$ of
\begin{eqnarray}
g(\epsilon,t_{\rm c})= 
I_{\rm amp}\left[
\frac{\epsilon}
{\sqrt{\epsilon^2+4t_{\rm c}^2}}
\right] {\rm tanh}\left[\frac{\sqrt{\epsilon^2+4t_{\rm c}^2}}{2k_{\rm B} T_{\rm e}}\right],
\label{eq:FermiHubbarddiCarlo}
\end{eqnarray}
the functional form expected for charge-based readout of a charge state anticrossing at finite temperature \cite{dicarlo2004}.
$I_{\rm amp}$ gives the amplitude of the current contrast, and $T_{e}$ is the effective electron temperature.  
State initialization primarily determines the ratio of the fit spin state populations, $P_{\rm S}/P_{\rm T}$.

While these formulae provide a concrete fit model, the splitting extracted from this technique is largely independent of model details. The midpoint of each branch
indicates the 
electrons' wave function is in an equal superposition of charge states, i.e., at an anticrossing. As long as this assumption holds, the distance between anticrossings will give $\Delta_{\rm SB}$ (with no upper-bound), regardless of higher-lying excited states, charge noise, or other effects that might distort the curves. By contrast, other parameters of Eq.~(\ref{eq:FermiHubbard3D}), in particular $t_{\rm c}$ and $T_e$, are much more sensitive to the precise fit model (see Supplemental Section V).  Although these model parameters give insight into the physics responsible for the measured lineshapes, in practice they only capture the slopes of the curves.

In the case of dot P1 where the two branches are well-separated, the energy-referred width of the SCC region in Fig.~\ref{fig2}a serves as a reasonably good estimate of $\Delta_{\rm SB}$, up to corrections of order $t_{\rm c}$. However, when SCC is not as robust, our technique is still effective at extracting $\Delta_{\rm SB}$. Figure~\ref{fig2}d shows data, taken from a similar device, in which the branches are not visibly distinct. Though the horizontal separation between branches only manifests as a widening of the combined curve near the anticrossings, we are still able to extract $\Delta_{\rm SB}$. To bound our confidence in such a fit, we consider the extreme case of trying to determine $\Delta_{\rm SB}$ from a single histogram (i.e. a vertical slice of a plot like Fig.~\ref{fig2}d). Using Eq.~(\ref{eq:FermiHubbarddiCarlo}) we can relate $\Delta_{\rm SB}$ to the separation of the histogram peaks, \hbox{$\eta\equiv g(\frac{\Delta_{\rm SB}}{2},t_c)-g(-\frac{\Delta_{\rm SB}}{2},t_c)$}. The remaining parameters of Eqs.~(\ref{eq:FermiHubbard3D}--\ref{eq:FermiHubbarddiCarlo}) can be fit to high precision, even at low splitting, so finding $\Delta_{\rm SB}$ reduces to a problem of fitting the separation between mixed Gaussian distributions with finite statistics.

Given $N$ single-shot measurements, we can calculate a confidence interval comparing an estimated $\widetilde{\Delta}_{\rm SB}$ to the actual ${\Delta}_{\rm SB}$. This calculation (performed in detail in Supplemental Section IV) gives
\begin{equation} \begin{split} \label{eq:finalerror}
\widetilde{\Delta}_{\rm SB} &\approx \Delta_{\rm SB} \pm  \Delta_{\rm SB} \frac{{\rm erf}^{-1}(C)}{\sqrt{N}}\left[1 + \frac{ \sigma_I^2}{\eta^2  P_{\rm S}P_{\rm T} }\right],\\
\eta &\approx \frac{I_{\rm amp}}{2t_c}\tanh{\left(\frac{t_c}{k_BT_{e}}\right)}\Delta_{\rm SB},
\end{split} \end{equation}
where ${\rm erf}^{-1}$ is the inverse error function and $C$ is the confidence of the estimate (e.g. ${\rm erf}^{-1}(.95)\approx 1.39$ for 95\% confidence). This uncertainty stems from shot noise, and exhibits the standard square root statistical improvement with number of shots. The first error term is the standard error in estimating the width of a distribution, while the second represents the added difficulty of measuring a separation between mixed Gaussians that is small compared to their individual variances. 
Though the splitting ($\Delta_{\rm SB}=28.1$ $\mu$eV) from the fit shown in Fig.~\ref{fig2}d is of the same order as both the tunnel coupling ($t_c=40.5$ $\mu$eV) and thermal broadening ($k_{\rm B}T_{\rm e}=8.6 ~\mu$eV), the last term in Eq.~(\ref{eq:finalerror}) gives an uncertainty of $\sim 2.5 ~\mu$eV due to histogram width (with 95\% confidence). Note that in practice, we never fit ${\Delta}_{\rm SB}$ to just one histogram; fitting to the full dataset---including an estimate of $\eta$ at each detuning---adds confidence to our estimate, so this uncertainty is an upper bound. Nonetheless, it shows that our technique can confidently measure splittings in real devices below 30 $\mu$eV, or lower with additional averaging.

\begin{figure}[h]
\includegraphics[width=1\linewidth]{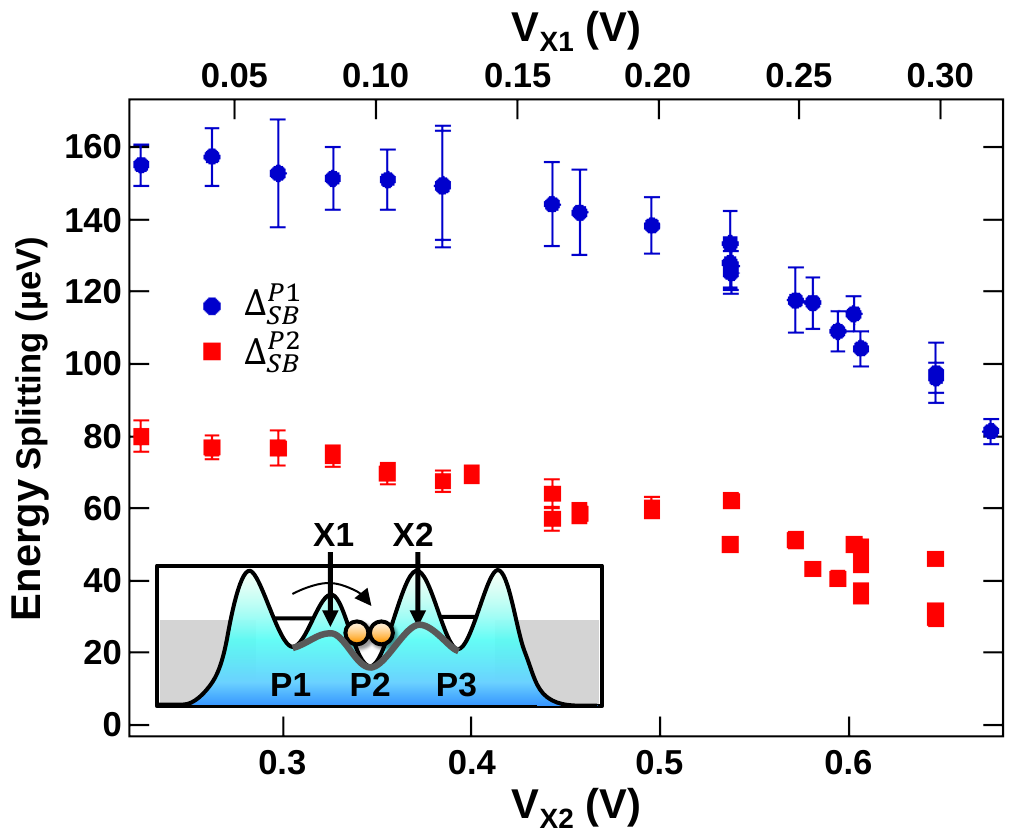}
\caption{Exchange gate modulation of two-electron excited-state energy.  Measured ground-to-excited state energy separation of dots $P1$ (blue) and $P2$ (red) as a function of voltage applied to neighboring exchange gates. To maintain spin-to-charge conversion, each $X2$ voltage (bottom axis) was compensated with an $X1$ voltage (top axis). Error bars: uncertainty calculated from Eqn. (\ref{eq:finalerror}) for $C$ = 0.95.}
\label{fig3}
\end{figure}

For dots with splittings greater than the tunnel coupling, the signal-to-noise ratio (SNR) of the singlet-triplet measurement is not fundamentally limited by $t_{\rm c}$ or $T_e$, but rather solely by $ \sigma_I$ and the current discrimination observed between the (1,1) and (2,0) charge states. Such is the case for dot P1, with an SNR of 6.5 near $\epsilon=\Delta_{\rm SB}/2$ (shown in Fig.~\ref{fig2}c).
 This limit is apparent by noting that at the detuning indicated by the dashed green line of Fig.~\ref{fig2}b, the two branches are fully (1,1) and (2,0) in character. The measured current contrast is dictated by the capacitive coupling from dots $P1$ and $P2$ to the charge sensor, a property of this particular gate design. The uncertainty in the measured current, $ \sigma_I$, of 7.6 pA is accounted for by the contributions of  $1/f$ charge noise (3.7 pA broadening for 5 $\mu$V/$\sqrt{\textrm Hz}$ at 1 Hz of gate-referred noise), Johnson noise (4.3 pA), HEMT input-referred noise (3.7 pA), and shot noise (3.6 pA) as described in Supplemental Section II.

A critical aspect of accurately resolving $\Delta_{\rm SB}$ using this technique is our ability to perform single-shot charge measurements at timescales much shorter than singlet-triplet $T_{1}$ processes. Since the measurement uncertainty scales quadratically with histogram width, $ \sigma_I$, the blurring of singlet and triplet histograms caused by $T_{1}$ decay during measurement is detrimental, especially when a small $\Delta_{\rm SB}$ and/or a high electron temperature limits SNR on fast timescales.   Near the region of SCC, we typically observe $T_{1}>100~\mu$s, while the total measurement integration times of steps (3) and (4) are 12.5 $\mu$s. This fast measurement time 
is accomplished by mounting cryogenic HEMTs near the device to reduce parasitic capacitance\cite{vink2004}, resulting in 60 dB of low-noise amplification up to 1 MHz.

\label{sec:VaryingOrb}

We now use this technique to measure how changes in confining potential affect $\Delta_{\rm SB}$ of dots $P1$ and $P2$. 
We sweep the bias voltages on gates $X1$ and $X2$ at a constant ratio,  as indicated by the bottom and top axes  of Fig.~\ref{fig3}, to deform the confinement  potentials while maintaining interdot tunnel couplings amenable to SCC. 
We observe that the energy splittings for each dot decrease with bias voltage by a factor of $\sim$2,  with maximum values of $\sim150$ $\mu$eV for $\Delta_{\rm SB}^{P1}$ and $\sim 80$ $\mu$eV for $\Delta_{\rm SB}^{P2}$.
We can rule out a continuous variation of the gate-dot lever arm as the source of this effect (see Supplemental Section I) and conclude that the underlying energy splitting is changing.

Identifying the physics responsible for this variation is challenging but important to improve device performance. 
 The relatively smooth, correlated change in $\Delta_{\rm SB}$ over a wide range of biases suggests that valley splitting alone is not responsible:   shifting wave functions would sample disorder at the interface of a SiGe alloy in a more random fashion \cite{Abadillo2018,Zhang2013}. Also unlikely is an  arrangement of steps that would  cause valley splittings to change with bias on distinct dots at nearly the same rate\cite{Friesen2007}.   Vertical electric fields easily tune valley splitting in MOS Si quantum dots  \cite{Yang2013,Gamble2016}, but the effect is much weaker in Si/SiGe, especially when controlled by neighboring barrier gates.  We therefore posit that in this case, the most likely  cause of the change in $\Delta_{\rm SB}$ with bias is a modification of the two-electron orbital confinement energy, i.e., the wave functions become more elliptical with forward bias on the barrier gates  \cite{Melnikov2006,Hanson2007}.  We note that valley-orbit hybridization is also possible and could be the cause of an apparent saturation in $\Delta_{\rm SB}$ at tighter confinements  \cite{Friesen2010}.   
Even though we believe that this device is limited by orbital splittings, we expect valley splittings to be similar in magnitude\cite{Neyens2018}.
The delicate interplay between these two degrees of freedom necessitates an $in~situ$ measurement technique as described here.

In summary, we have described an excited-state spectroscopy method for measuring the two-electron ground-to-excited state energy separation of quantum dots. The measured energy splitting is  directly relevant to state preparation and measurement of singlet/triplet based qubits.   This technique, which is only suitable in systems where charge measurements are possible within spin $T_{1}$,  is applied to quantum dots tuned for spin-to-charge conversion.  
In showing the tunability of $\Delta_{\rm SB}$ with confinement gate bias, we demonstrate the utility of this technique in optimizing device performance.   This measurement technique should be useful to all quantum dot architectures that require robust spin-blockade.

The authors gratefully acknowledge T. Ladd, M. Gyure, M. Reed, A. Pan, D. Kim for helpful discussions.

\end{document}


\title{Supplemental Material:  Spin-Blockade Spectroscopy of Si/SiGe Quantum Dots}
\maketitle

\section{Lever Arm Strength} \label{sec:leverarm}
Crucial to the accurate extraction of the ground-to-excited state splitting, $\Delta_{\rm SB}$, using the technique described in the main text,  is 
the conversion of detuning from the swept bias voltage to the actual difference in the chemical potential energies of the respective dots.
 While several methods for doing this have been presented elsewhere (cf. photon assisted tunneling\cite{Stoof1996,Oosterkamp1998}), we outline our two-step process here. 

\begin{figure}[h]
\includegraphics[width=0.7\linewidth]{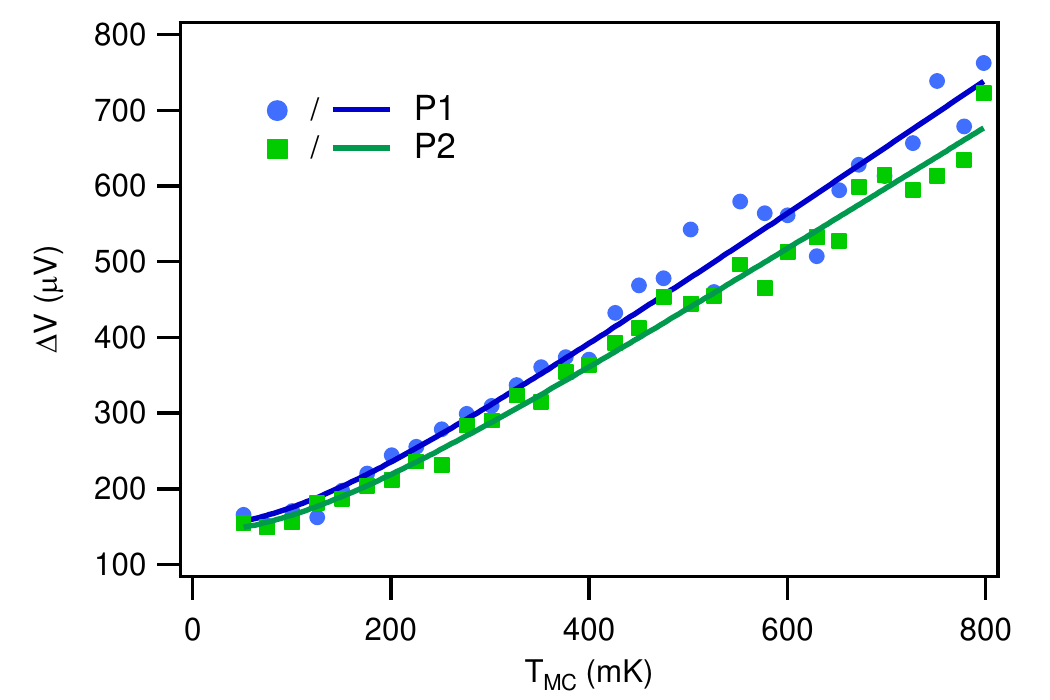} 
\caption{Charge transition width vs. refrigerator base temperature.  Measured width, $\Delta V({T}_{\rm MC})$, of the $P1$ (1,0)-(2,0) (blue circles) and $P2$ (3,0)-(3,1) (green squares) charge transitions as a function of refrigerator base temperature, $T_{\rm MC}$. Blue and green lines show fits to corresponding data sets to extract lever arm strength.}
\label{fig2}
\end{figure}

First, we obtain the lever arm strength, $\alpha$, of each plunger gate by fitting a charge transition width as a function of refrigerator temperature. For the $P1$ plunger gate we monitor the (1,0)-(2,0) charge transition. For the $P2$ plunger gate we monitor the (3,0)-(3,1) charge transition. The measured linewidth of each transition as a function of refrigerator base temperature is shown in Fig.~\ref{fig2}. These curves are fit to a phenomenological expression which assumes that contributions to the linewidth stem from an incoherent sum of the measured temperature of the mixing chamber, $T_{\rm MC}$, and a constant electronic `effective' temperature, $T_{\rm e}$, as 
\begin{equation}
\Delta V(T_{\rm MC})=\frac{k_{B}~\sqrt[]{T_{\rm MC}^{2}+T_{\rm e}^{2}}}{\alpha}.
\label{eq:AlphaTempFcn}
\end{equation}
Fits to the data (blue and green lines in Fig.~\ref{fig2}) yield $\alpha_{P1}=0.095$ eV/V and $\alpha_{P2}=0.104$ eV/V. The difference between $\alpha_{ P1}$ and $\alpha_{P2}$ (9\%) 
is much lower than the fitting error and better reflects the variation found within an ensemble of lever arm measurements. 
The resulting uncertainty in the ratio of detuning biases to energy directly affects the accuracy with which we can extract $\Delta_{\rm SB}$.

Second, we measure the cross capacitance of gate ${ P2}$ on an electron loaded under ${P1}$ compared to the capacitance of gate ${ P1}$ on the same electron, which we denote $G_{P1,P2}$, by measuring the slope of a $P1$ loading line with respect to bias  $V_{P2}$.   $G_{ P2,P1}$ can be measured similarly.
The four measured factors $\alpha_{P1}$, $\alpha_{ P2}$, $G_{ P1, P2}$, and $G_{P2,P1}$, 
together with the relative bias swept on gates $P2$ and $P1$ ($\Delta V_{ P2}/\Delta V_{ P1}$ where $\Delta V_{ P1}-\Delta V_{ P2}$ is the detuning bias), allow us to convert the $P1$ voltage range supporting spin-blockade, $\Delta V_{\rm SB}$, to energy:
\begin{equation}
\Delta_{\rm SB}=\big|{\alpha_{P1}-\alpha_{P2}\cdot G_{ P2,P1}-\frac{\Delta V_{P2}}{\Delta V_{P1}}(\alpha_{P2}-\alpha_{P1}\cdot G_{ P1,P2})\big|} \Delta V_{\rm SB}.
\label{eq:ScaleFactor}
\end{equation}

One effect which may give rise to the variation in $\Delta_{\rm SB}$ with bias discussed in the main text is a varying voltage-to-energy scaling caused by changes in cross capacitance or lever arm strength, $\alpha$.  
We account for changes in cross capacitance by updating the values of $G_{P1,P2}$ and $G_{P2,P1}$ over the course of the sweep in Fig. 3 of the main text.
Although changes in the electrostatic potential can certainly affect $\alpha$ if the electron is localized far from under the center of the corresponding gate, we find this interpretation of the data of main text Figure 3 to be inconsistent with our observations. 
Measurements of $\alpha$ from devices similar to the one studied here are typically performed in a low-exchange gate biasing regime (e.g. near the left-hand side of Fig. 3). By comparison, the measurements shown here were performed with the $X1$ and $X2$ exchange gates forward biased ($V_{X2}$ = 0.675 V and $V_{X1}$ = 0.318~V, right-hand side of Fig. 3). 
The agreement between typical values of lever arm strength ($\sim$0.1 eV/V) with those measured here contrasts with the factor of two change in $\Delta_{\rm SB}$ observed in the main text. 
Therefore, we conclude that the large tunability of $\Delta_{\rm SB}$ is not confounded by significant changes in lever arm strength.

\section{Cryogenic Readout Electronics}

\begin{figure}[h]
\includegraphics[width=0.7\linewidth]{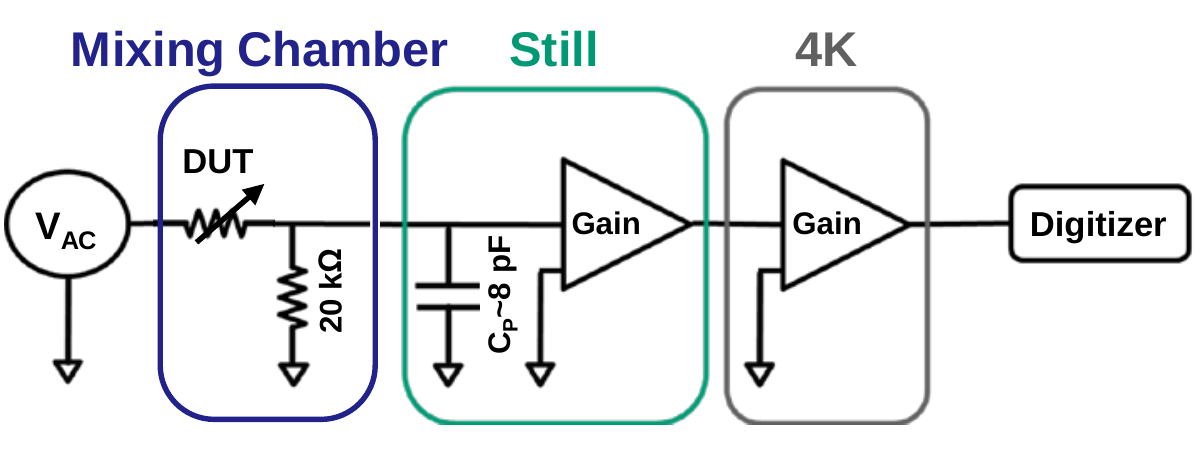}
\caption{Experiment measurement chain.  At the output of the device under test (DUT), a 20~kOhm shunt resistor on the mixing chamber stage converts current to voltage. With $\sim$8~pF of parasitic capacitance, this arrives at the first HEMT amplification stage held at $\sim 1$ K before passing through the second stage held at $\sim 4$ K, providing a gain of 60 dB up to 1 MHz.}
\label{MsrmntChain}
\end{figure}

Measuring capacitive shifts at the charge sensor requires a chain of cryogenic and room temperature electronics leading to both the source and drain electrodes. Changes in charge sensor current are detected using  lock-in amplification techniques. The lock-in tone is a series of alternating square pulses from an arbitrary waveform generator at room temperature sent down the source line.  This line has 10,000:1 voltage division 
at room temperature to mitigate voltage noise and is filtered by a commercial low pass filter with a 120 MHz cutoff frequency. Alternating current driven through the charge sensor is converted back into voltage using a $\sim$20~kOhm shunt resistor, which  reduces the impedance of the high impedance charge sensor ($\sim$100~kOhm).    
Since the measured voltage scales linearly with the value of the shunt resistor, we increase our bandwidth by minimizing parasitic capacitance between the drain and the amplifiers, rather than by lowering the resistance. By mounting the pHEMTs on standoffs located near the device \cite{vink2004}, we achieve a capacitance of  $\sim$8~pF, allowing for a measurement bandwidth of 1 MHz, well above the $1/f$  corner frequency of our electronics. The voltage signal is amplified $\sim$60~dB using 2 stages of amplification (Avago ATF-38143 pHEMTs) held at $\sim 1$ K and $\sim 4$ K. The amplified voltage is delivered using rf coaxial cables  and digitized at room temperature using commerical ADCs  where the lock-in tone through the charge sensor is demodulated and converted into units of conductance. Components of this measurement chain that contribute to the total noise floor of the measurement include the Johnson noise of the shunt resistor (332 pV/$\sqrt{\rm Hz}$ at 100 mK), shot noise (253 pV/$\sqrt{\rm Hz}$), and the input-referred noise of the pHEMT amplifiers ($\sim 250$ pV/$\sqrt{\rm Hz}$). Integrating these sources with our measurement filter function and adding the result to the $1/f$  charge noise contribution (3.65 pA broadening for 5 $\mu$V/$\sqrt{\rm Hz} ~$at$~1~{\rm Hz}$ of gate-referred noise), we obtain a total broadening of 7.7 pA, in excellent agreement with the observed value of 7.6 pA from Fig. 2c of the main text.

\section{Fermi-Hubbard Model Fitting}
\label{sec:FermiHubbard}
We derive our fit model, Eqs.~(1--3) in the main text, from a Fermi-Hubbard model, following \cite{dicarlo2004}. Near the charge boundaries that support spin-to-charge conversion (SCC), the system's ground state is a superposition of (1,1) and (2,0) charge states, connected by a two-electron tunnel coupling, $t_{\rm c}$, with detuning $\epsilon$. We assume that the electrons have a valley/orbital degree of freedom with excitation energy $\Delta_{\rm{}SB}$, so that the ground $\ket{g}_{20}$ and excited $\ket{e}_{20}$ states of two-electrons occupying the same dot (the (2,0) charge configuration indicated by subscript) have splitting $\Delta_{\rm{}SB}$.  At detunings near the region of SCC, the system's ground state is dominated by the lowest allowed energy states:  $\ket{g}_{11}$ and $\ket{g}_{02}$ for spin-singlets and $\ket{g}_{11}$ and $\ket{e}_{02}$ for the spin-triplet.   $\ket{g}_{02}$ is forbidden for spin-triplets by the Pauli exclusion priniple.
In this reduced basis, we can write the Hamiltonian as $H=H_{\rm S} + H_{\rm T}$ where
\begin{align}
H_{\rm{}S}&\approx t_{\rm c}^{\rm{}S}\big(\ket{g}_{11}\prescript{}{20}{\bra{g}}+{\rm h.c.}\big) - \epsilon\ket{g}_{20}\prescript{}{20}{\bra{g}},\label{eq:singletH}\\
H_{\rm{T}}&\approx t_{\rm c}^{\rm{}T}\big(\ket{g}_{11}\prescript{}{20}{\bra{e}}+{\rm h.c.}\big) - (\epsilon-\Delta_{\rm SB})\ket{e}_{20}\prescript{}{20}{\bra{e}},\label{eq:tripletH}
\end{align}
and the  singlet and triplet $t_{\rm c}$ may in principle be different, especially when the excited state is determined by the valley degree of freedom. 
In this case, $t_{\rm c}^{\rm T}/t_{\rm c}^{\rm S}$ is a function of the ``valley mixing angle", the rotation from one dot to the other of the relative phase between  superpositions of the valley states \cite{Culcer2010,Tagliaferri2018}.  
The Hamiltonians of Eqs.~(\ref{eq:singletH}--\ref{eq:tripletH}) are identical up to an offset in $\epsilon$ and potentially different tunnel couplings, so without loss of generality we focus on $H_{\rm{}S}$ in the following analysis.

For a given $\epsilon$, $H_{\rm{}S}$ has an energy gap of \hbox{$E(\epsilon, t_{\rm c})=\sqrt{\epsilon^2+4t_{\rm c}^2}$}, and its ground state has (2,0) population \hbox{$P_{20}=\left(1+\epsilon/E(\epsilon, t_c)\right)/2$}. At finite temperature $T_e$, we will find the system in its excited state with probability ${\rm exp}(-E(\epsilon, t_c)/k_BT_e)$. Weighting $P_{20}$ by this value we obtain
\begin{equation}
P_{20}=\frac{1}{2}\left[1+\frac{\epsilon}{E(\epsilon, t_c)}\right]\tanh\left[\frac{E(\epsilon,t_c)}{2k_BT_e}\right].
\end{equation}
Combining this with parameters describing the charge sensor, which we assume exhibits a linear response, we arrive at Eq.~(3) of the main text.

If we expand our model by including additional excited states or broadening mechanisms, $P_{20}$ will still be $\sim50\%$ at the (1,1)--(2,0) anticrossing, and $\Delta_{\rm SB}$ can be accurately extracted as the distance between the midpoints of the singlet and triplet branches. On the other hand, extracted values of $t_{\rm c}$ (or $T_{\rm e}$) can depend strongly on such changes. As a vivid example, we consider a process that dephases our system in the charge basis, possibly due to charge noise in $\epsilon$ or backaction from the measurement itself.  We consider the density matrix describing our system, $\rho$, as it evolves under the coherent Hamiltonian $H_{\rm{}S}$ as well as Lindbladians describing thermalization and charge dephasing. Denoting the ground and excited eigenstates of $H_{\rm{}S}$ as $\ket{-}$ and $\ket{+}$ --- not to be confused with the $\ket{g}_{ c}$ and $\ket{e}_c$, the ground and excited states of a given charge configuration $c$ --- the master equation governing these dynamics is
\begin{equation} \label{eq:mastereq} \begin{split}
\dot{\rho}(t)=&-i\left[H_{\rm{}S}, \rho(t)\right]+\mathcal{L}_{\rm thermal}(\rho(t))+\mathcal{L}_{\rm charge}(\rho(t)),\\
\mathcal{L}_{\rm thermal}(\rho) =& -\gamma\left(\bra{+}\rho\ket{+}-\bra{-}\rho\ket{-}-\tanh\left[\frac{E(\epsilon,t_{\rm c})}{2k_BT_{\rm e}}\right]\right)\big(\ket{+}\bra{+}\:-\:\ket{-}\bra{-}\big)\\
&-\frac{\gamma}{2}\big(\bra{+}\rho\ket{-}\ket{+}\bra{-} + {\rm h.c.}),\\
\mathcal{L}_{\rm charge}(\rho) =& -\frac{\kappa}{2}(\prescript{}{11}{\bra{g}}\rho\ket{g}_{20}\ket{g}_{11}\prescript{}{20}{\bra{g}} + {\rm h.c.}),\\
\end{split} \end{equation}
where $\gamma$ and $\kappa$ are the thermalization and charge dephasing rates, respectively.

\begin{figure}[h]
\includegraphics[width=0.7\linewidth]{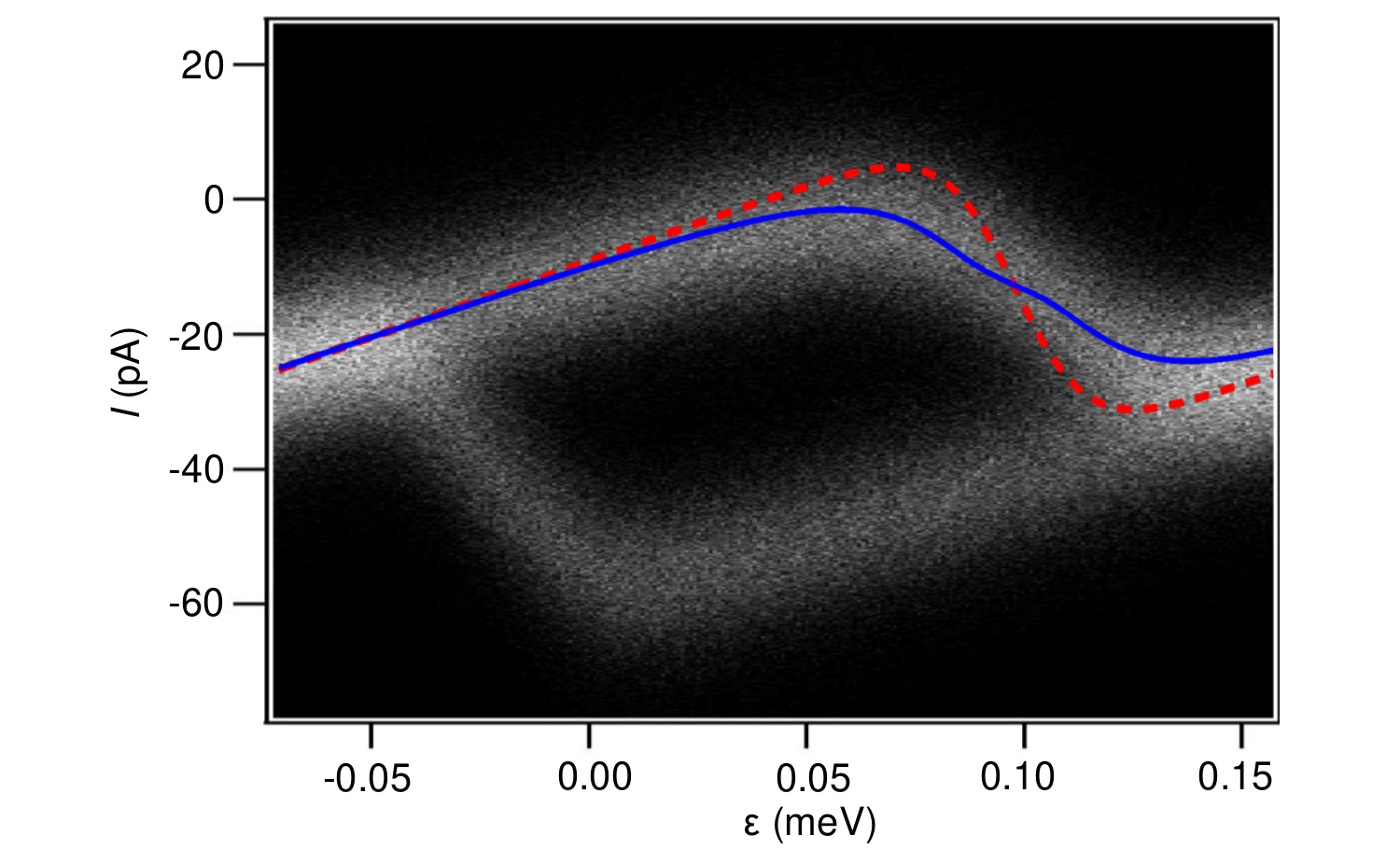} 
\caption{Curve fits using $t_{\rm c}$ determined by spin funnel. With no charge dephasing (red dashed line), $t_{\rm c}$ is inadequate to match the observed broadening. With charge dephasing at $150\times$ the rate of thermalization (blue solid line), the broadening is recovered.}
\label{fig:chargeDephasing}
\end{figure}

The quantity of interest in our measurement technique is the steady state probability $P_{20}^{\rm ss}\equiv\lim_{t\rightarrow\infty} P_{20}(t)=\lim_{t\rightarrow\infty}\prescript{}{20}{\bra{g}}\rho (t)\ket{g}_{20}$.  Numerically, we find that charge dephasing can lead to large, multiplicative broadening of $P_{20}^{\rm ss}$ as a function of $\epsilon$.  The degree of broadening depends on $\kappa/\gamma$, i.e. the ratio of charge- to thermal-decoherence rates. If charge dephasing is much faster than thermalization, this can increase the extracted $t_{\rm c}$ by two orders of magnitude or more. This may explain the large discrepancy in values of $t_{\rm c}$ extracted from fitting to Eqs.~(1--3) of the main text (assuming fixed $T_{\rm e}$) compared to other established techniques (see Supplemental Section~\ref{sec:TunnelCoupling}); a $\kappa$ of $\sim150\gamma$ can resolve the discrepancy (see Fig.~\ref{fig:chargeDephasing}). However, including charge-state dephasing in the fit leaves $T_{\rm e}$, $t_{\rm c}$, $\kappa$, and $\gamma$ as highly correlated parameters, and our fits are not sufficiently well-resolved to distinguish them. It is therefore too difficult to determine the impact of charge dephasing without an independent measurement of its magnitude,
and so  we mainly suggest it as an illustration of how broadening mechanisms can affect parameter estimation in our technique.

\section{Resolving Small $\Delta_{\rm SB}$}

We now derive rough bounds on the minimum splitting that can be resolved with the technique outlined in the main text.  Distinguishing the singlet and triplet branches is most difficult when they are similar in shape and close together, so we assume $t_{\rm c}^{\rm S}\approx t_{\rm c}^{\rm T}$ and $\Delta_{\rm SB}<t_{\rm c}$. In this limit, we can determine the vertical separation of the branches, $\eta$, by a linear expansion of the fit function. At $\epsilon-\epsilon^0=\Delta_{\rm SB}/2$, where separation between the branches is near its maximum,
\begin{equation} \label{eq:iform}
\eta \approx \frac{I_{\rm amp}}{2t_{\rm c}}\tanh\left(\frac{t_{\rm c}}{{\rm k_B}T_{\rm e}}\right) \Delta_{\rm SB} \equiv \lambda \Delta_{\rm SB},
\end{equation}
where we have introduced the slope $\lambda$ to condense notation. If the singlet and triplet histograms have variance $\sigma_I^2$ and means separated by $\eta$, the combined histogram's total variance is given by
\begin{equation} \label{eq:sigmaform}
\sigma_{\rm tot}^2=\sigma_I^2+P_{\rm S}P_{\rm T}\eta^2.
\end{equation}
We can combine Eqs.~(\ref{eq:iform}) and (\ref{eq:sigmaform}) to solve for $\Delta_{\rm SB}$ in terms of $\sigma_{\rm tot}^2$.

The precision of any estimate of $\sigma_{\rm tot}^2$ is bounded by finite statistics. After $N$ measurements, we can find an approximate $\widetilde{\sigma}_{\rm tot}^2$ which bounds the actual $\sigma_{\rm tot}^2$ according to
\begin{equation} \label{eq:errorform}
\widetilde{\sigma}_{\rm tot}^2 = \sigma_{\rm tot}^2\left(1 \pm \frac{2{\rm erf}^{-1}(C)}{\sqrt{N}}\right),
\end{equation}
where ${\rm erf}^{-1}$ is the inverse error function and $C$ is the confidence of the estimate. Writing versions of Eq.~(\ref{eq:sigmaform}) for both actual and estimated variables and plugging the results into Eq.~(\ref{eq:errorform}), we find a relation between the actual $\eta$ and estimated $\widetilde{\eta}$ which simplifies to
\begin{equation} \label{eq:finalerror}
\widetilde{\eta} = \eta \sqrt{1 \pm \frac{2{\rm erf}^{-1}(C)}{\sqrt{N}} \left(1 + \frac{\sigma_I^2}{\eta^2 P_{\rm S}P_{\rm T}}\right)}.
\end{equation}
The second error term implies $N$ must scale with $(\sigma_I/\eta)^4$ to maintain a fixed uncertainty; this determines the minimum $\eta$ that can be extracted with a reasonable number of measurements.

If the uncertainty in Eq.~(\ref{eq:finalerror}) is not overly large, we may wish to combine this with uncertainty in the other fit parameters to bound our overall confidence in $\Delta_{\rm SB}$. Let $\delta X$ denote our fractional uncertainty in parameter $X$, i.e. $\widetilde{X}=X(1\pm\delta X)$. Eq.~(\ref{eq:errorform}) can be understood as an expression of $\delta\sigma_{\rm tot}$. Expanding Eqs.~(\ref{eq:iform}) and (\ref{eq:sigmaform}) to first order in each $\delta$, and assuming that errors add quadratically, we find
\begin{align}
&\delta\Delta_{\rm SB}^2 \lesssim \delta\eta^2 + \delta \lambda^2,\\
&\delta\eta^2 \approx \left(\frac{{\rm erf}^{-1}(C)}{\sqrt{N}} \left(1 + \frac{\sigma_I^2}{\eta^2 P_{\rm S}P_{\rm T}}\right)\right)^2 + \left(\frac{|P_{\rm S} - P_{\rm T}|}{2P_{\rm T}} \delta P_{\rm S}\right)^2 + \left(\frac{\sigma_I^2}{\eta^2P_{\rm S}P_{\rm T}}\delta\sigma_I\right)^2. \label{eq:alltheerrors}
\end{align}
$\delta \lambda$ can be expanded into a function of $\delta t_{\rm c}$, $\delta I_{\rm amp}$, and $\delta T_{\rm e}$, but these are generally correlated fit parameters whose errors may not add independently, so we express their uncertainty collectively. If we assume that all fit parameters other than $\Delta_{\rm SB}$ are known to high precision and keep only the shot noise term of Eq.~(\ref{eq:alltheerrors}), we recover Eq.~(4) of the main text. Finally, it is worth emphasizing that all of the error sources discussed in this section are exclusive to small-$\Delta_{\rm SB}$ fits. If $\Delta_{\rm SB}$ is large enough to resolve the singlet and triplet branches, the fit is determined by horizontal distance rather than histogram width, and uncertainty is dominated by our measurement of the lever-arm, as described in Sec.~\ref{sec:leverarm}.

\section{Measuring $\Delta_{\rm SB}$ of the inner dot}

\begin{figure}[h]
\includegraphics[width=0.9\linewidth]{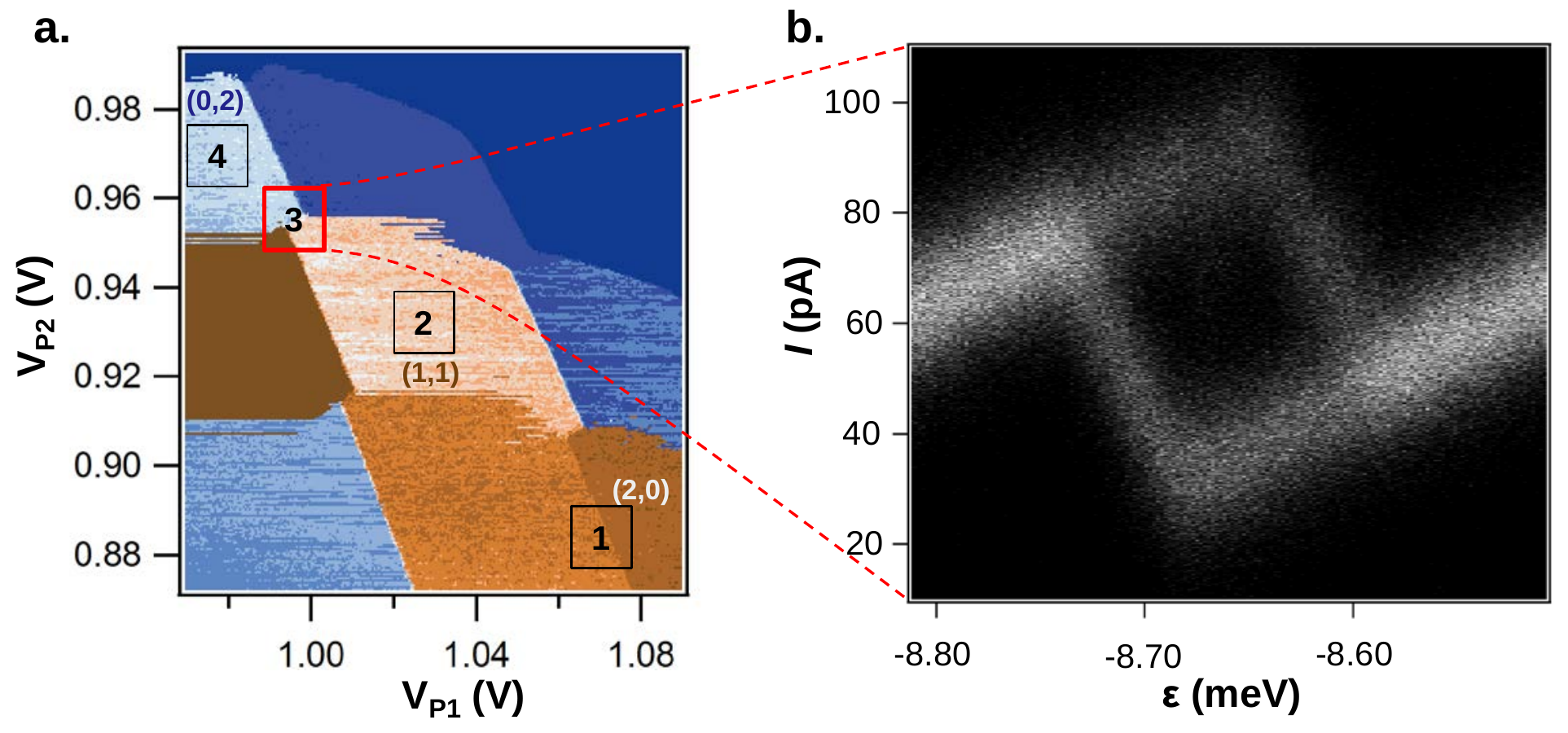}
\caption{{\bf{a}}, Slight modifications in the pulse sequence are needed to extract $\Delta_{\rm SB}$ of the inner dot ($P2$). The singlet initialization (1) and dephasing (2) steps remain the same, while measurement of spin blockade (3) is performed at the other charge-conserving boundary, (0,2)-(1,1), with the reference measurement point (4) set within the (0,2) charge cell. {\bf{b}}, Although measurements of the inner dot share qualitative similarities with measurements of the outer $P1$ dot, the spin blockade region for $P2$ is at a negative detuning point ($\epsilon\approx-8.6$~meV). }
\label{P2DotData}
\end{figure}

As mentioned in the main text, our excited-state spectroscopy technique can be performed at any charge boundary supporting spin blockade. Since  the `inner' dot $P2$ is more isolated from the bath, state preparation must be performed on an `outer' dot (i.e. $P1$). Thus when measuring $\Delta_{\rm SB}^{{ P}2}$, the same (1) spin preparation and (2) dephasing steps occur as when measuring $\Delta_{\rm SB}^{{ P}1}$, followed by measurement near the region of SCC at the (1,1)-(0,2) charge boundary, with the final charge reference measurement set within the (0,2) charge cell, as shown for (3) and (4) in Fig.~\ref{P2DotData}a. These slight modifications of the pulse sequence allow $\Delta_{\rm SB}$ to be meaured on two quantum dots without changes to the electrostatic configuration.

\label{sec:P2DotData}

\section{Measurement of Tunnel Coupling}
\label{sec:TunnelCoupling}

\begin{figure}[h]
\includegraphics[width=0.7\linewidth]{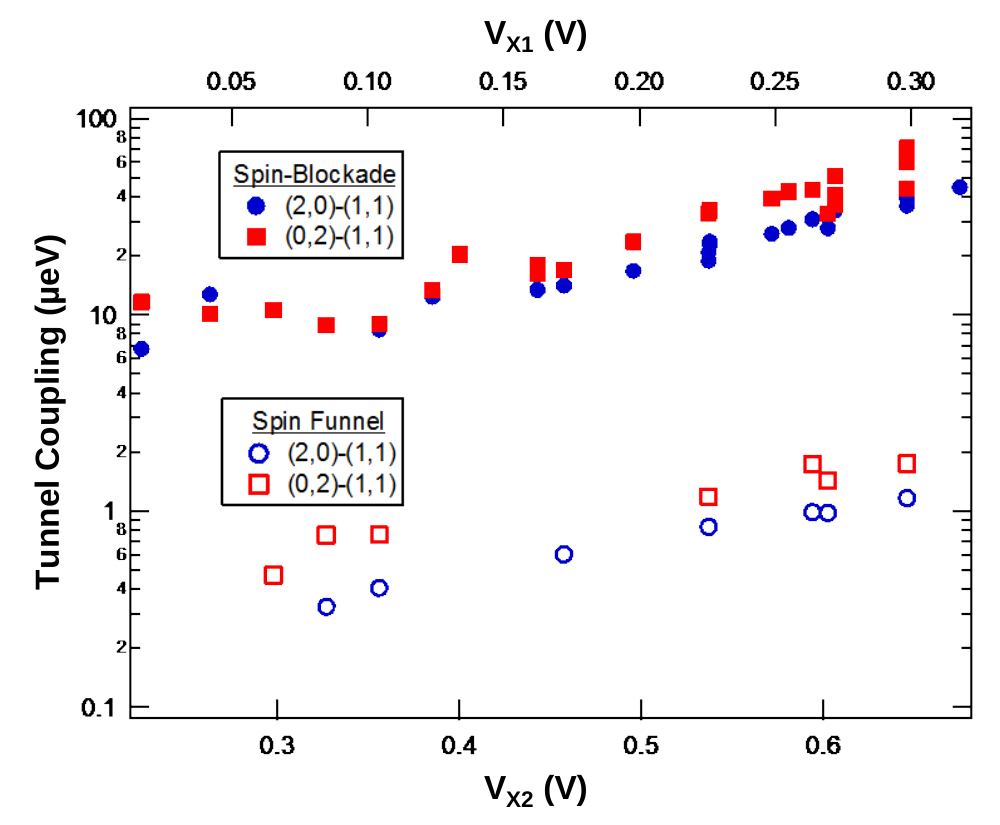}
\caption{Tunnel couplings extracted from spin-blockade fits compared to spin funnel measurements.  Tunnel coupling at the (2,0)-(1,1) anticrossing (blue) and (0,2)-(1,1) anticrossing (red), extracted from a Fermi-Hubbard model fit to single-shot data using fit funcitons in main text at fixed $T_{\rm e}$ (solid markers) and tunnel couplings from spin funnel measurements (open markers).}
\label{TunnelCoup}
\end{figure}

In addition to the two-electron ground-to-excited state splitting, $\Delta_{\rm SB}$, fitting data to Eq.~(1) of the main text yields parameters describing the tunnel coupling, $t_{\rm c}$.   The solid markers of Figure~\ref{TunnelCoup} show  $t_{\rm c}$ extracted from the data of Fig.~3 of the main text, fit assuming a fixed value of $T_{\rm e}$. This is  contrasted with $t_{\rm c}$ extracted from the more established spin funnel measurement technique, shown by open markers \cite{Petta2005,Petta2008,maune2012}. Notably, $t_{\rm c}$ extracted from spin-funnel measurements (which also consistently agree with Landau-Zener estimates of tunnel coupling, not shown here) are more than 10$\times$ smaller than those estimated using the fit technique described in the main text. 
Even at the highest $X2$ voltages, where tunnel coupling dominates any temperature terms in the fit formula, the discrepancy persists. 
As noted in the main text, any effect that broadens the curves when fit to Eqs. (1--3) will lead to an overestimation of $t_{\rm c}$, assuming that $T_{\rm e}$ has been reliably measured some other way. One such effect is the interplay of incoherent processes --- charge state dephasing and eigenstate $T_{1}$ --- during the measurement process, as discussed in Supplemental Section~\ref{sec:FermiHubbard}.   This suggests that this relatively simple, site-based coherent Fermi-Hubbard interaction model used to derive the fit functions in the main text is insufficient to describe real experimental data.

\begin{figure}[h]
\includegraphics[width=0.7\linewidth]{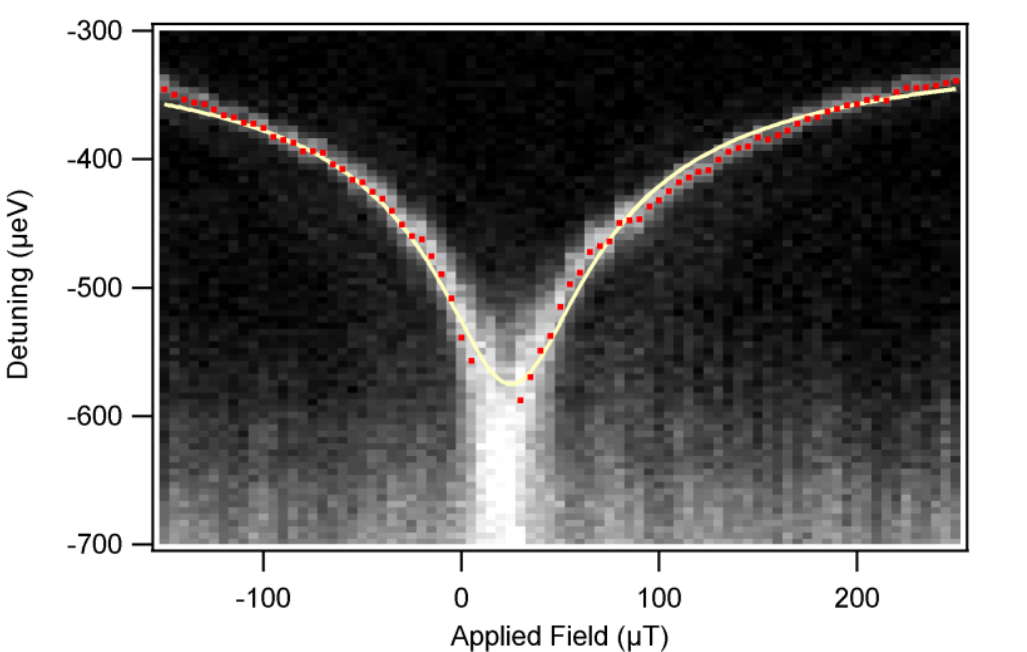}
\caption{Spin funnel measurement of tunnel coupling.  Grayscale contrast highlights the probability of measuring a Pauli-blockaded spin state as a function of detuning and applied magnetic field. Red markers denote extracted peak position for each column. The yellow line shows a fit to the red markers using Eq.~(\ref{eq:SpinFunnelFcn}).}
\label{SpinFunFig}
\end{figure}

The `spin funnel' measurements of the (2,0)-(1,1) and (0,2)-(1,1) tunnel couplings shown in Fig.~\ref{TunnelCoup} were performed by measuring the location of the $S_{0}-T_{\pm}$ avoided crossing as a function of applied magnetic field. This established measurement technique offers a reliable comparison for the tunnel coupling extracted other ways. The sequence employed first prepares a singlet state, biases near the anticrossing of interest for a reference measurement, and then pulses to varying negative detunings (toward the (1,1) cell center) with a dwell time of 10 $\mu$s, after which the final pulse returns to the spin-to-charge readout window for measurement. The 10 $\mu$s dwell time is chosen to be long enough to allow spin-orbit and hyperfine interactions to induce mixing between $S_{0}$ and $T_{\pm}$ states at their anticrossing. This yields an increased probability of transitioning into the lower of the $T_{-}$ and $T_{+}$ states at the detuning of that anticrossing (see Fig.~\ref{SpinFunFig}).

The location of the $S_{0}-T_{\pm}$ anticrossing is extracted by fitting a Gaussian profile to each column. The resulting peak location vs. magnetic field is then fit to the functional form:
\begin{equation}
\varepsilon = y_{0}-\frac{t_{c}^{2}}{g\mu_{B}~\sqrt[]{(B-B_{0})^{2}+B_{\bot}^{2}}},
\label{eq:SpinFunnelFcn}
\end{equation}
from which we extract the tunnel coupling, $t_{c}=1.07 \pm0.02~\mu eV$, as well as the vertical ($B_{0}=25~\mu T$) and horizontal ($B_{\bot}=36~\mu T$) offset fields.